% Date: 1999 June

% Version 6: Vanilla TeX after JK return from Pasadena
% Version 7: Vanilla TeX with small changes by LCH
% Version 8: Vanilla TeX with small changes by LCH after referee's report 
%            (10/20/99)
%
\input psfig.tex

\magnification = \magstephalf
\hsize=16.1truecm  \hoffset=0.6truecm  \vsize=23.2truecm  \voffset=-0.1truecm

\def\dblbaselines{\baselineskip=15pt \lineskip=0pt \lineskiplimit=0pt}

\def\vs{\vskip 8pt} \def\vss{\vskip 6pt} \def\vsl{\vskip \baselineskip}
\parskip = 0pt % default is \parskip = 0pt+1pt
\nopagenumbers

    \font\bbf=cmbx12

\def\makeheadline{\vbox to 0pt{\vskip-30pt\line{\vbox to8.5pt{}\the
                               \headline}\vss}\nointerlineskip}
\def\toppageno{\headline={\hss\tenrm\folio\hss}}
\def\footnoterule{\kern-3pt \hrule width \hsize \kern 2.6pt \vskip 3pt}
\def\omit#1{\empty}
\pretolerance=15000  \tolerance=15000
\def\ts{\thinspace}  \def\cl{\centerline}
          \def\b{\kern -0.2em}
\def\0{\phantom{0}}    \def\00{$\phantom{000000}$}

\def\1{\phantom{1}}  
\def\etal{{\it et~al.~}} 
\def\gapprox{$_>\atop{^\sim}$}  \def\lapprox{$_<\atop{^\sim}$}
\newdimen\sa  \def\sd{\sa=.1em  \ifmmode $\rlap{.}$''$\kern -\sa$
                                \else \rlap{.}$''$\kern -\sa\fi}
              \def\se{\sa=.1em  \rlap{.}{''}\kern -\sa}
              \def\dgd{\sa=.1em \ifmmode $\rlap{.}$^\circ$\kern -\sa$
                                \else \rlap{.}$^\circ$\kern -\sa\fi}
\newdimen\sb  \def\md{\sa=.06em \ifmmode $\rlap{.}$'$\kern -\sa$
                                \else \rlap{.}$'$\kern -\sa\fi}
\def\kms{km~s$^{-1}$}
\def\s{\ifmmode ^{\prime\prime} \else $^{\prime\prime}$ \fi}
\def\min{\ifmmode ^{\prime} \else $^{\prime}$ \fi}
\def\deg{\ifmmode ^{\circ} \else $^{\circ}$ \fi}
\def\msun {M$_{\odot}$~}  \def\msund{M$_{\odot}$}  
\def\mbh{$M_{\bullet}$}   
\def\lsund{$L_\odot$}

% Luis Ho's TeX definitions:

\def\e#1{$\times$10$^{#1}$}
\def\lum{erg s$^{-1}$}
\def\al{$\alpha$}
\def\lamb{$\lambda$}

\def\pp{\parshape 2 0truein 6.5truein .3truein 6.2truein}

%  \pp is a simple definition to define a paragraph shape in 
%      which the first line is not indented, but subsequent lines are.
%      suitable for references and figure captions.
\def\apjo{{\it Astrophys. J. }}
\def\aplo{{\it Astrophys. J. Lett. }}
\def\nato{{\it{Nature }}}
\def\mnraso{{\it{M.N.R.A.S. }}}
\def\annrevo{{\it Ann. Rev. Astr. Astrophys. }}
\def\paspo{{\it Pub. Astr. Soc. Pac. }}

\parindent=20pt

\dblbaselines

\cl{{\bbf SUPERMASSIVE BLACK HOLES IN ACTIVE GALACTIC NUCLEI}\footnote{$^1$\b}
{To appear in {\it Encyclopedia of Astronomy and Astrophysics}}}

\vskip 0.5truecm

\cl{Luis C. Ho\footnote{$^2$\b}{Carnegie Observatories, 813 Santa Barbara 
St., Pasadena, CA 91101-1292}{\ts}
and John Kormendy\footnote{$^3$\b}{Department of Astronomy, RLM 15.308,
University of Texas, Austin, TX 78712-1083}}

\vskip 0.5truecm

\cl{1.~SUPERMASSIVE BLACK HOLES AND THE AGN PARADIGM}
\vs

      Quasars are among the most energetic objects in the Universe.  We now know
that they live at the centers of galaxies and that they are the most dramatic
manifestation of the more general phenomenon of active galactic nuclei (AGNs).
These include a wide variety of exotica such as Seyfert galaxies, radio 
galaxies, and BL Lacertae objects.  Since the discovery of quasars in 1963, much
effort has gone into understanding their energy source.  The suite of proposed
ideas has ranged from the relatively prosaic, such as bursts of star formation
that make multiple supernova explosions, to the decidedly more colorful, such as
supermassive stars, giant pulsars or ``spinars,'' and supermassive black holes
(hereinafter BHs).  Over time, BHs have gained the widest acceptance.  The 
key observations that led to this consensus are as follows.

     Quasars have prodigious luminosities. Not uncommonly, $L \sim 10^{46}$ erg
s$^{-1}$; this is 10 times the luminosity of the brightest galaxies.  Yet they
are tiny, because they vary on timescales of hours.  From the beginning, the 
need for an extremely compact and efficient engine could hardly have been more
apparent.  Gravity was implicated, because collapse to a black hole is the most
efficient energy source known.  The most cogent argument is due to Donald 
Lynden-Bell (1969, {\it Nature}, {\bf 223}, 690).  He showed that any attempt 
to power quasars by nuclear reactions alone is implausible.  First, a lower 
limit to the total energy output of a quasar is the energy, $\sim 10^{61}$ 
erg, that is stored in its radio-emitting plasma halo.  This energy weighs 
$10^{40}$ g or $10^7$ \msund.  But nuclear reactions produce energy with an 
efficiency of only $\epsilon$ = 0.7\ts\%.  Then the waste mass left behind in 
powering quasars would be at least $M_\bullet \simeq 10^9$ \msund. Lynden-Bell 
argued further that quasar engines are $2R$ \lapprox \ts10$^{15}$ cm in 
diameter because large variations in quasar luminosities are observed on 
timescales as short as 10 h.  But the gravitational potential energy of 
$10^9$ \msun compressed into a volume as small as 10 light hours is 
$\sim G M_\bullet^2 / R$ \gapprox \ts10$^{62}$ erg.  As Lynden-Bell noted, 
``Evidently although our aim was to produce a model based on nuclear fuel, we 
have ended up with a model which has produced more than enough energy by 
gravitational contraction.  The nuclear fuel has ended as an irrelevance.''  
We now know that the total energy output is larger than the energy that is 
stored in a quasar's radio source; this strengthens the argument.  Meanwhile, 
a caveat has appeared: the objects that vary most rapidly are now thought to 
contain relativistic jets that are beamed at us.  This boosts the power of 
a possibly small part of the quasar engine and weakens the argument that the 
object cannot vary on timescales less than the light travel time across it.  
But this phenomenon would not occur at all if relativistic motions were not 
involved, so BH-like potential wells are still implicated. These 
considerations suggest that quasar power derives from gravity.

      The presence of deep gravitational potentials has long been inferred from
the large velocity widths of the emission lines seen in optical and ultraviolet
spectra of AGNs.  These are typically 2000 to 10,000 km s$^{-1}$.  If the large
Doppler shifts arise from gravitationally bound gas, then the binding objects 
are both massive and compact.  The obstacle to secure interpretation has always
been the realization that gas is easy to push around: explosions and ejection of
gas are common astrophysical phenomena.  The observation that unambiguously 
points to relativistically deep gravitational potential wells is the detection
of radio jets with plasma knots that are seen to move faster than the speed of
light, $c$.  Apparent expansion rates of 1 -- 10 $c$ are easily achieved if the 
true expansion rate approaches $c$ and the jet is pointed almost at us.

\pageno=2 \toppageno

      The final pillar on which the BH paradigm is based is the observation 
that many AGN jets are well collimated and straight.  Evidently AGN engines can
remember ejection directions with precision for up to $10^7$ yr.  The natural 
explanation is a single rotating body that acts as a stable gyroscope.  
Alternative AGN engines that are made of many bodies -- like stars and
supernovae -- do not easily make straight jets.  

      A variety of other evidence also is consistent with the BH picture, but
the above arguments were the ones that persuaded a majority of the astronomical 
community to take the extreme step of adopting BHs as the probable engine for
AGN activity.  In the meantime, BH alternatives such as single supermassive 
stars and spinars were shown to be dynamically unstable and hence short-lived. 
Even if such objects can form, they are believed to collapse to BHs.

      The above picture became paradigm long before there was direct evidence 
for BHs.  Dynamical evidence is the subject of the present and following 
articles.  Meanwhile, there are new kinds of observations that point directly to
BH engines.  In particular, recent observations by the {\it Advanced Satellite
for Cosmology and Astrophysics\/} ({\it ASCA\/}) have provided strong evidence
for relativistic motions in AGNs.  The X-ray spectra of many Seyfert galaxy 
nuclei contain iron K$\alpha$ emission lines (rest energies of 6.4 -- 6.9 keV;
see Figure 1). These lines show enormous Doppler broadening --- in some cases 
approaching 100,000 km s$^{-1}$ or 0.3$c$ --- as well as asymmetric line 
profiles that are consistent with relativistic boosting and dimming in the 
approaching and receding parts, respectively, of BH accretion disks as small 
as a few Schwarzschild radii. 

      The foregoing discussion applies to the most powerful members of the AGN
family, namely quasars and high-luminosity Seyfert and radio galaxies.  It is 
less compelling for the more abundant low-luminous objects, where energy 
requirements are less demanding and where long jets or superluminal motions are 
seen less frequently or less clearly.  Therefore a small but vocal competing 
school of thought continues to argue that stellar processes alone, particularly 
those that occur during bursts of star formation, can reproduce many AGN 
characteristics.  Nonetheless, dynamical evidence suggests that BHs {\it do\/}
lurk in some mildly active nuclei, and, as discussed in the next article, even 
in the majority of inactive galaxies. 

\vfill\eject

%\hskip 0.3truein
%\psfig{file=nandra.ps,height=3.0truein,angle=0}

\cl{\null} \vskip 2.8truein

\includegraphics{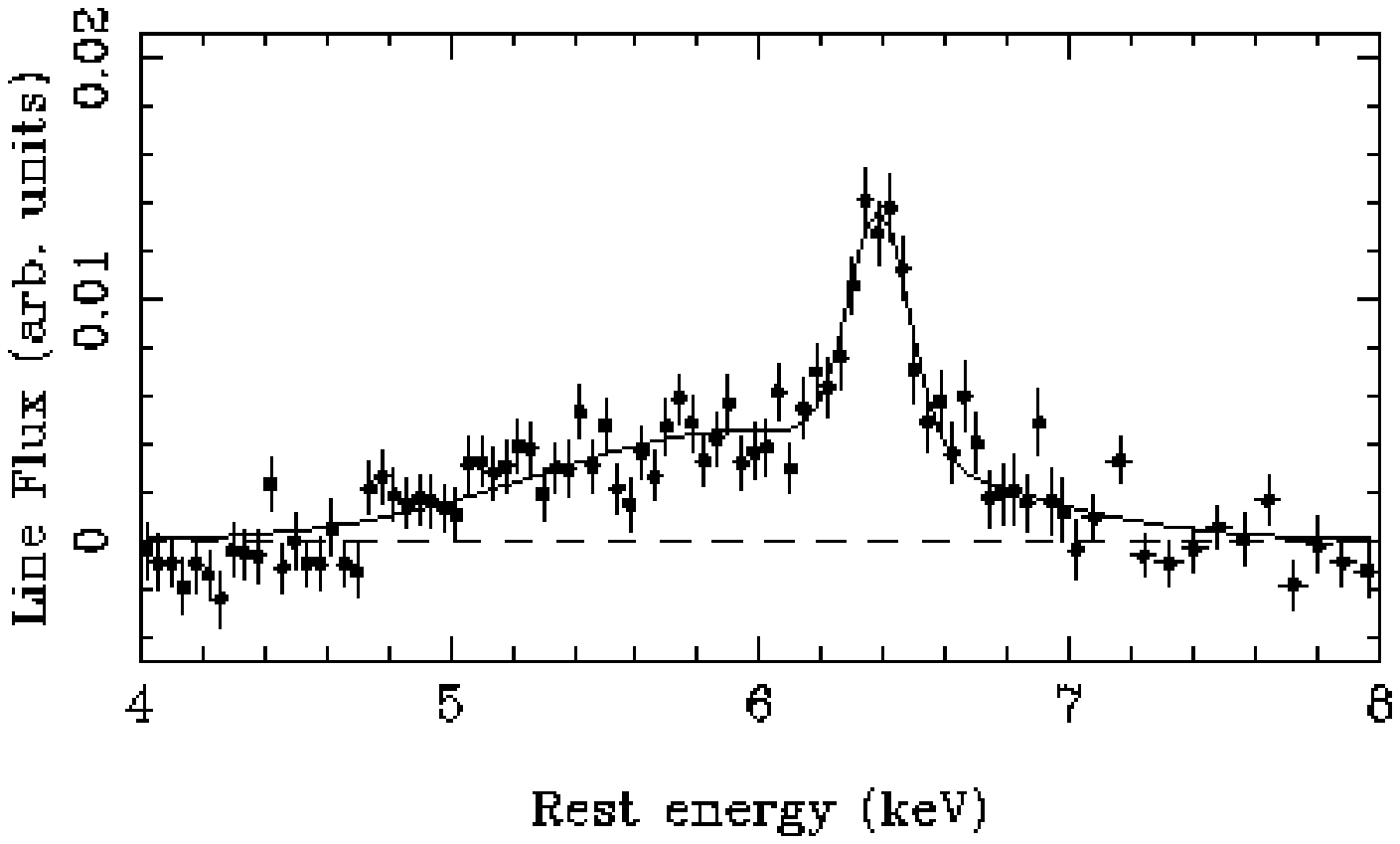}

{\bf Figure 1.} A composite x-ray spectrum of Seyfert nuclei taken with
{\it ASCA} showing the relativistically broadened Fe K$\alpha$ line.  The
solid line is a fit to the line profile using two Gaussians, a narrow
component centered at 6.4 keV and a much broader, redshifted component.
[Figure adapted from Nandra, K., \etal \apjo {\bf 477}, 602 (1997).]
%\vfill\eject

\vs 
\cl{2.~MEASURING AGN MASSES: DIRECT METHODS}
\vs

      Very general arguments suggest that quasar engines have masses \mbh\ 
$\sim$ $10^6$ to $10^9$ \msund.  Gravitational collapse is believed to liberate
energy with an efficiency of $\epsilon \simeq 0.1$; Lynden-Bell's arguments then
imply that typical remnant masses are \mbh\ $\sim$ $10^8$ \msund.  Better 
estimates can be derived by asking what we need in order to power quasar 
luminosities, which range from 10$^{44}$ to 10$^{47}$ erg s$^{-1}$ or 
10$^{11}$ to 10$^{14}$ \lsund.  For $\epsilon$ = 0.1, the engine must consume 
0.02 to 20 \msund\ yr$^{-1}$.  How much waste mass accumulates depends on how
long quasars live.  This is poorly known.  If they live long enough to make
radio jets that are collimated over several Mpc, and if their lifetimes are
conservatively estimated as the light travel time along the jets, then quasars
last \gapprox 10$^7$ yr and reach masses \mbh\ \gapprox 10$^5$ to 10$^8$ \msund.
But the most rigorous lower limit on \mbh\ follows from the condition that the 
outward force of radiation pressure on accreting matter not overwhelm the inward
gravitational attraction of the engine, a condition which, admittedly, 
strictly holds only if the accreting material and the radiation have spherical 
symmetry.  This so-called Eddington limit requires that 
$L\,\leq\,L_{\rm E}\, \equiv\,(4\pi G c m_p/\sigma_T) M_\bullet$ =
1.3\e{38} ($M_\bullet$/\msund) \lum, or equivalently that $M_\bullet\,\geq$ 
8\ts\e{5} ($L$/10$^{44}$ \lum) \msund.  Here $G$ is the gravitational constant, 
$m_p$ is the mass of the proton, and $\sigma_T$ is the Thompson cross section 
for electron scattering.  We conclude that we are looking for BHs with masses
\mbh\ $\sim$ 10$^6$ to 10$^9$ \msund.  Finding them has become one of the
``Holy Grails'' of astronomy because of the importance of confirming or
disproving the AGN paradigm.

      AGNs provide the impetus to look for BHs, but active galaxies are the 
most challenging hunting ground.  Stellar dynamical searches first found central
dark objects in inactive galaxies (see the next article), but they cannot be 
applied in very active galaxies, because the nonthermal nucleus outshines the 
star light.  We can estimate masses using the kinematics of gas, but only if it 
is unperturbed by nongravitational forces.  Fortunately, this complication can
be ruled out {\it a posteriori\/} if we observe that the gas is in Keplerian 
rotation around the center, i.{\ts}e., if its rotation velocity as a function of
radius is $V(r) \propto r^{-1/2}$.  We can also stack the cards in our favor by
targeting galaxies that are only weakly active and that appear to show gas disks
in images taken through narrow bandpasses centered on prominent emission lines.

\vfill
 
% The following is a smaller version of the figure in case you need it.
%\special{psfile=M87Disk_editted-small.ps hoffset=40 voffset=-30
%                                         hscale=55  vscale=55}

\includegraphics{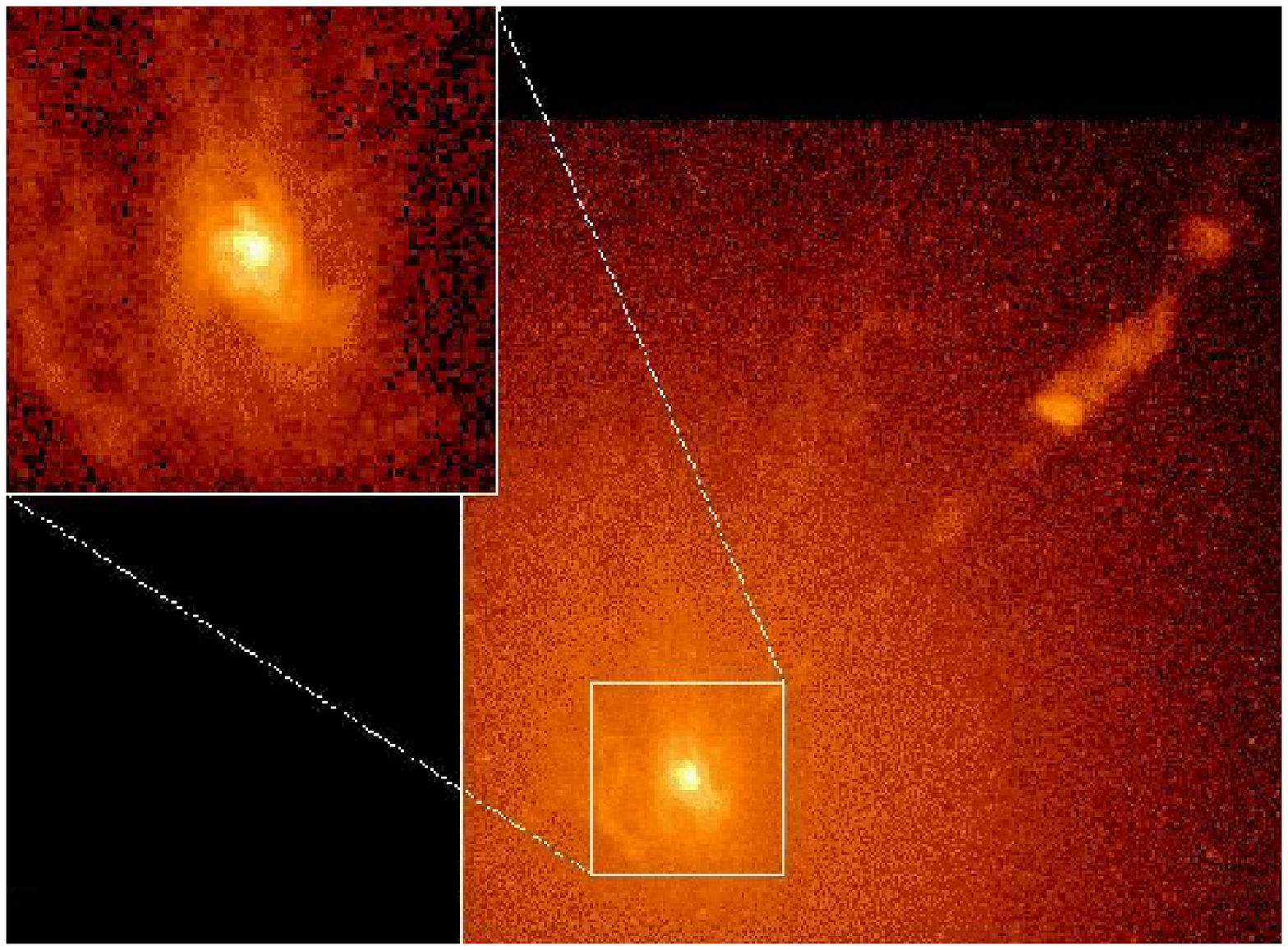}
 
{\bf Figure 2a.} {\it HST\/} image of the ionized gas disk near the center of
the giant elliptical galaxy M{\ts}87.  The data were taken with the Second Wide
Field/Planetary Camera through a filter that isolates the optical emission
lines H\al\ and [N~II] \lamb\lamb6548, 6583.  The left inset is an expanded viewof the gas disk; for an adopted distance of 16.8 Mpc, the region shown is
5\s\ $\times$ 5\s\ or 410 $\times$ 410 pc. The disk has a major axis
diameter of $\sim$ 150 pc, and it is oriented perpendicular to the optical jet.
[Image courtesy of NASA/Space Telescope Science Institute, based on data
originally published by Ford, H.~C., \etal \apjo {\bf 435}, L27 (1994).]
 
\vsl\vsl\vsl\vsl
 
\eject
 
\cl{\null}\vfill
 
\includegraphics{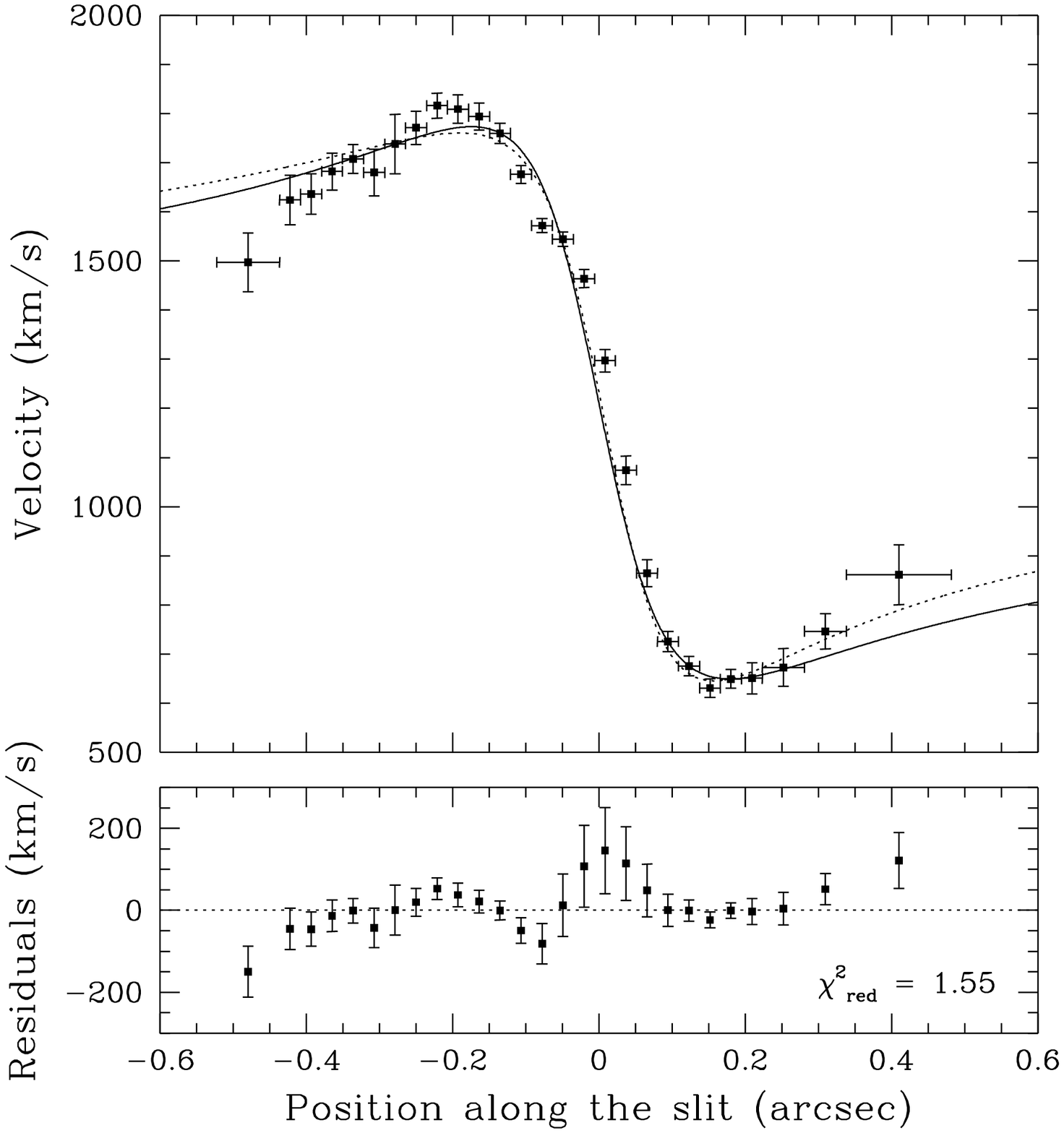}
 
{\bf Figure 2b.} Optical emission-line rotation curve for the nuclear disk in
M{\ts}87.  The data were taken with the Faint Object Camera on {\it HST}.  The
curves in the upper panel correspond to two different Keplerian thin disk
models, and the bottom panel shows the residuals for the best-fitting model.
[Figure adapted from Macchetto, F., \etal \apjo {\bf 489}, 579 (1997).]
 
\vsl\vsl
 
\eject
 
\cl{\null}\vfill
 
\includegraphics{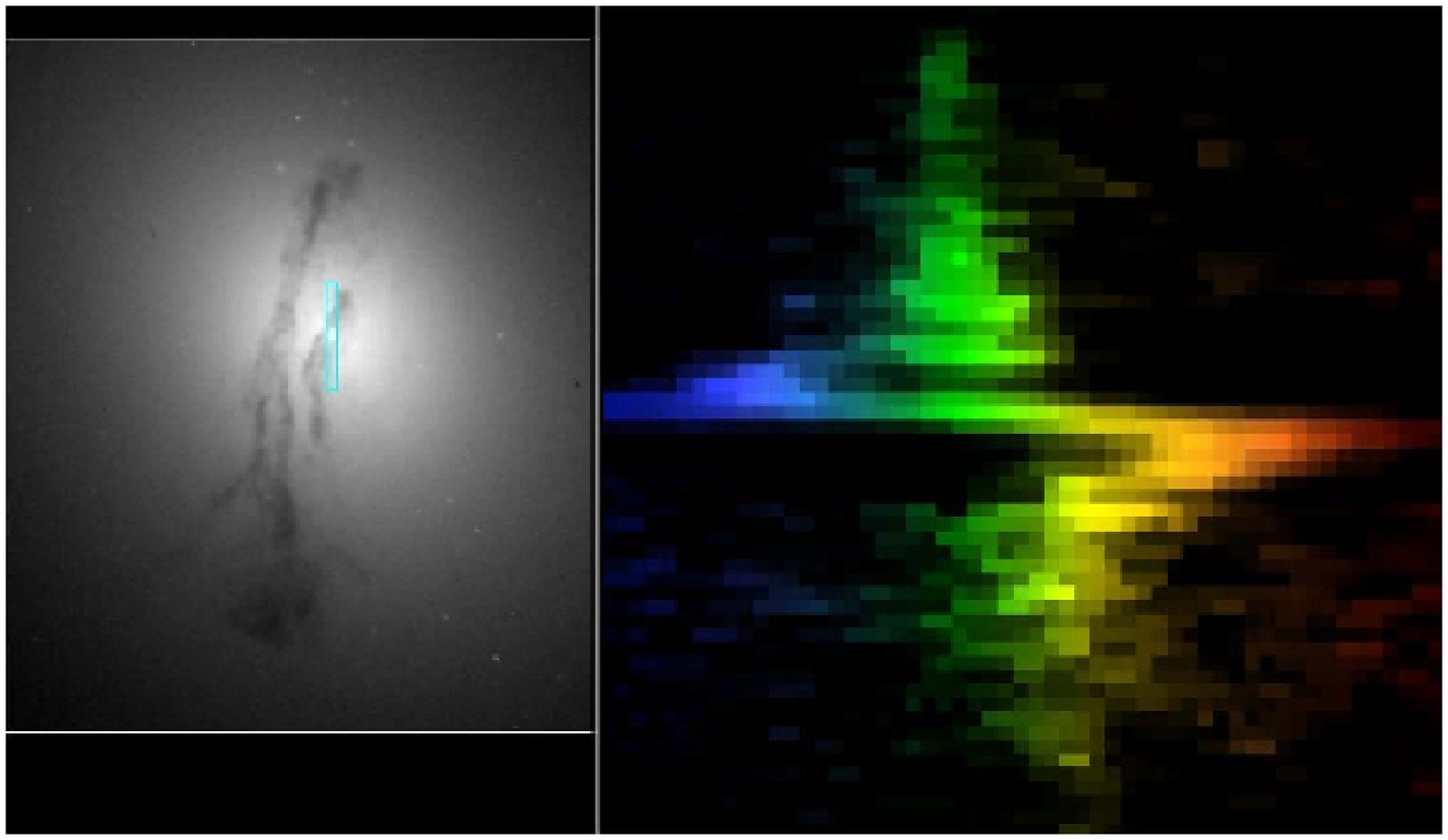}
 
{\bf Figure 3.} ({\it Left\/}) {\it HST\/} image of the central region of the
giant elliptical galaxy M{\ts}84; the box measures 22\s\ $\times$ 19\s\ or
1.8 kpc $\times$ 1.6 kpc for an adopted distance of 16.8 Mpc.  The data were
taken with the Second Wide Field/Planetary Camera through a filter that
isolates the optical emission lines H\al\ and [N~II] \lamb\lamb6548, 6583.
The slit of the Space Telescope Imaging Spectrograph was placed along the
major axis of the nuclear gas disk (blue rectangle).  ({\it Right\/})
Resulting spectrum of the central 3\s\ (240 pc).  The abscissa is velocity and
the ordinate is distance along the major axis.  The spectrum shows the
characteristic kinematic signature of a rotating disk.  The velocity scale is
coded such that blue and red correspond to blue and red shifts, respectively;
the total velocity range is 1445 \kms.  [Image courtesy of NASA/Space
Telescope Science Institute, based on data originally published by Bower,
G.~A., \etal \aplo {\bf 483}, L33 (1997) and Bower, G.~A., \etal \aplo
{\bf 492}, L111 (1998).]
 
\vsl\vsl
 
\eject

\vs
\cl{2.1 Kinematics of Optical Emission Lines}
\vs

    High-resolution optical images taken with ground-based telescopes and
especially with the {\it Hubble Space Telescope\/} ({\it HST\/}) show that many
giant elliptical galaxies contain nuclear disks of dust and ionized gas.  The 
most famous case is M{\ts}87 (Figure 2a).  The disk measures $\sim$ 150 pc
across, and its rotation axis is closely aligned with the optical and radio jet.
This is in accord with the BH accretion picture.  The disk is in Keplerian 
rotation (Figure 2b) around an object of mass \mbh\ $\simeq$ 3\ts\e{9} \msund.
Furthermore, this object is dark: the measured mass-to-light ratio exceeds 100
in solar units, and this is much larger than that of any known population of 
stars.  Moreover, the dark mass must be very compact: the velocity field 
limits its radial extent to be less than 5 pc.  Therefore its density exceeds 
10$^7$ \msund\ pc$^{-3}$.  Another illustration of this technique is given in 
Figure 3.  M{\ts}84, also a denizen of the Virgo cluster of galaxies, is a 
twin of M{\ts}87 in size, and it, too, harbors an inclined nuclear gas disk 
(diameter $\sim$80 pc), whose rotation about the center betrays an invisible 
mass of \mbh\ $\simeq$ 2\e{9} \msund. Other cases are reported (NGC 4261, 
NGC 6251, NGC 7052), and searches for more are in progress.

\vs
\cl {2.2 Kinematics of Radio Masers}
\vs

      A related approach exploits the few cases where 22 GHz microwave maser 
emission from water molecules has been found in edge-on nuclear disks of gas.
Particularly strong ``megamasers'' allow radio astronomers to use interferometry
to map the velocity field with exquisite angular resolution.  In the most 
dramatic application of this method, the Very Long Baseline Array was used to
achieve resolution 0\sd0006 -- 100 times better than that delivered by {\it HST}
-- in observations of the Seyfert galaxy NGC 4258.  This is only 6 Mpc away, so
the linear resolution was a remarkable 0.017 pc. The masers trace out a slightly
warped annulus with an inner radius of 0.13 pc, an outer radius of 0.26 pc, and
a thickness of $<$ 0.003 pc (Figure 4, {\it left\/}).  The masers with nearly 
zero velocity with respect to the galaxy are on the near side of the disk along
the line of sight to the center, while the features with high negative 
(approaching) and positive (receding) velocities come from the disk on either 
side of the center.  High velocities imply that 3.6\ts\e{7} \msund\ of binding
matter resides interior to $r$ = 0.13 pc. What is most compelling about NGC 4258
is the observation that the rotation curve is so precisely Keplerian (Figure 4,
{\it right\/}).  From this result, one can show that the radius of the mass
distribution must be $r$ \lapprox 0.012 pc. If the central mass were {\it not\/}
a BH, its density would be extraordinarily high, $\rho$ $>$ 5\ts\e{12} \msund\
pc$^{-3}$.  This is comparable to the density of the dark mass at the center of
our Galaxy (see following article).  Under these extreme conditions, one can 
show that a cluster of stellar remnants (white dwarf stars, neutron stars, and
stellar-size black holes) or substellar objects (planets and brown dwarfs) are 
short-lived.  Astrophysically, these are the most plausible alternatives to a 
BH.  Therefore the dynamical case for a supermassive black hole is stronger in
NGC 4258 and in our Galaxy than in any other object.

\eject

\cl{\null}\vfill

%%%\special{psfile=miyoshi_fig2.eps hoffset=-40 voffset=70 
%%%                                 hscale=45  vscale=45}

%%%\special{psfile=miyoshi_fig3.eps hoffset=205 voffset=80 

\includegraphics{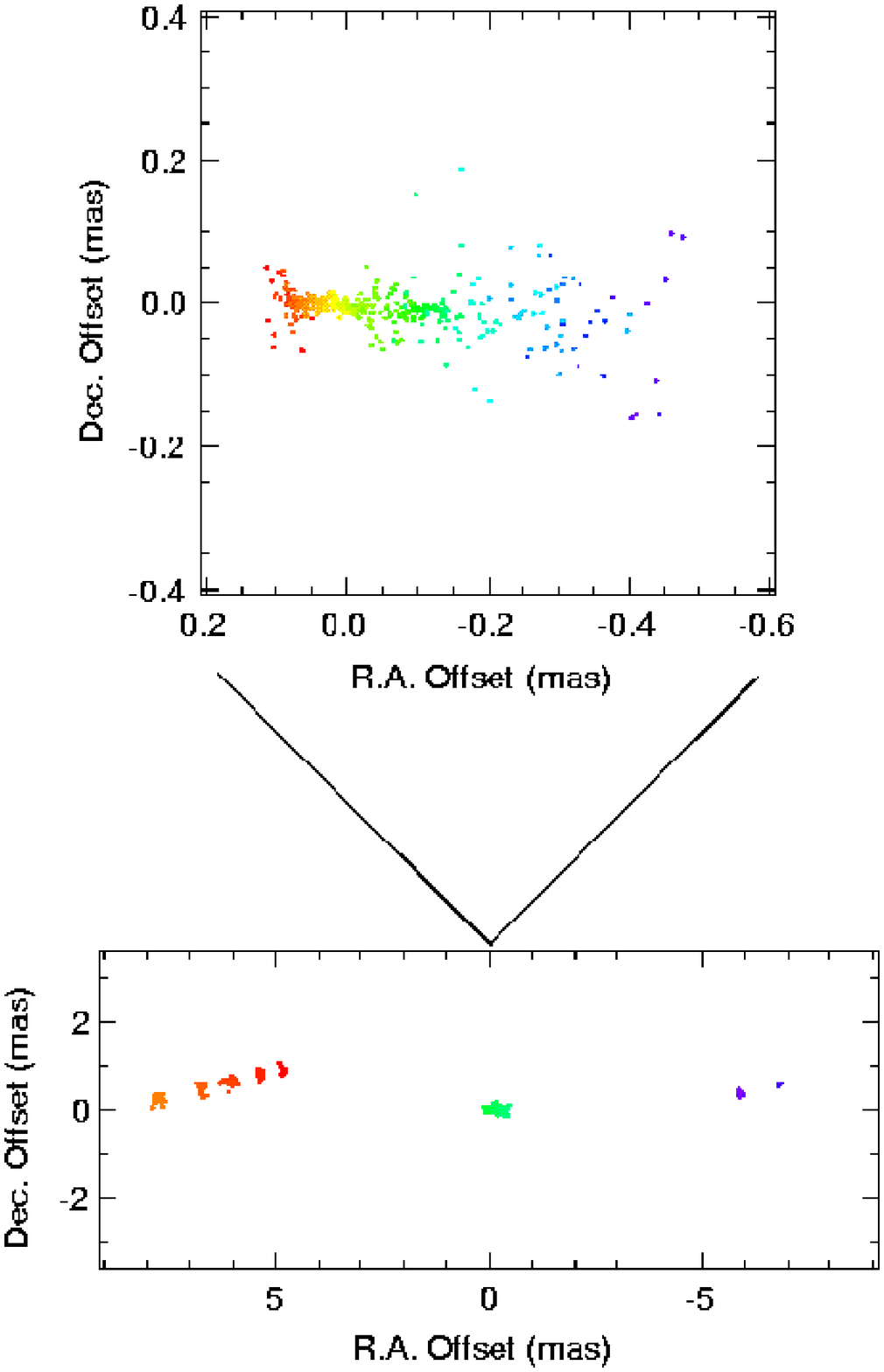}

\includegraphics{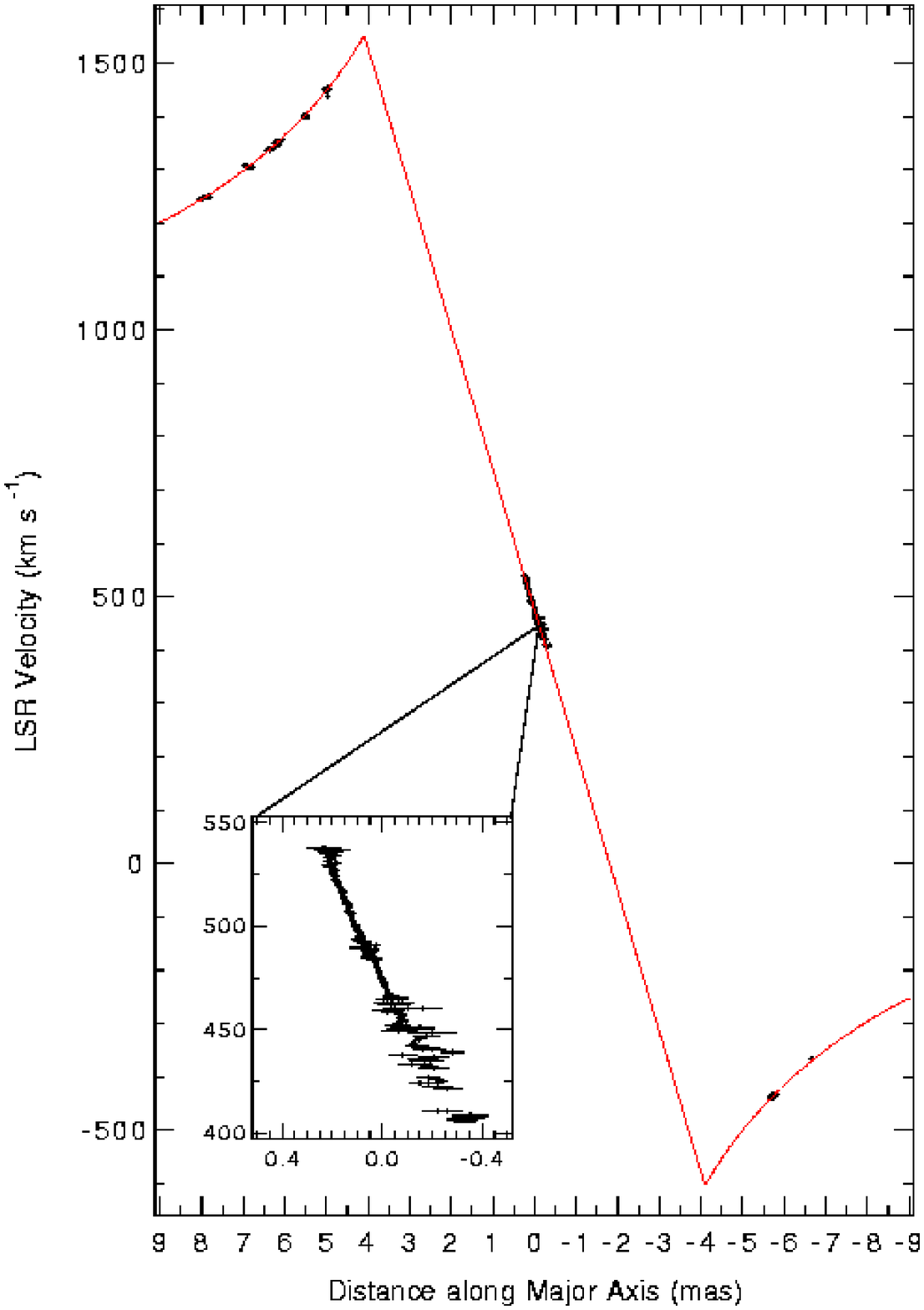}

{\bf Figure 4.} ({\it Left\/}) Spatial distribution of the water masers in NGC 
4258, color-coded so that blue and red correspond to blueshifted and redshifted
velocities, respectively.  The maser spots are distributed in a thin, warped 
annulus that is only 4$^\circ$ from edge-on. For an adopted distance of 6.4 Mpc,
1 mas = 0.031 pc.  The top panel shows an expanded view of the emission near the
systemic velocity of the galaxy.  ({\it Right\/})  Light-of-sight velocity as a
function of distance along the major axis of the annulus.  The high-velocity 
features are accurately fitted by a Keplerian model, overplotted as a continuous
line.  The emission near the systemic velocity, magnified in the inset, lies at
nearly constant radius in the front part of the disk along the line of sight to
the center.  The linear velocity gradient results from the change in projection
of the rotation velocity.
[Figure adapted from Miyoshi, M., \etal {\it Nature\/} {\bf 373}, 127 (1995).]

\eject

\cl {3.~MEASURING AGN MASSES: INDIRECT METHODS}
\vs

      Direct dynamical measurements are impractical for more luminous and more
distant AGNs.  The tremendous glare from the nucleus outshines the circumnuclear
emission from stars, and the violent conditions near the center are
likely to subject the gas to nongravitational forces.  Indirect methods of 
estimating central masses have been devised to provide a reality check for these
more difficult objects.

\vs
\cl {3.1 Fitting the Spectra of Accretion Disks}
\vs

      As material falls toward a black hole, it is believed to settle into an
accretion disk in which angular momentum is dissipated by viscosity.  From the 
virial theorem, half of the gravitational potential energy $U$ is radiated.
Therefore the luminosity is
$$
L\,=\,{1\over2}\,{dU\over{dt}}\,=\,{1\over2}\,{G M_\bullet 
                                        \dot{M_\bullet}\over{r}}~. \eqno{(1)}
$$
At sufficiently high accretion rates $\dot{M_\bullet}$, the gas is optically
thick, and the disk radiates as a thermal blackbody:
$$
      L\,=\,2 \pi r^2 \sigma T^4~.                                 \eqno{(2)}
$$
Here $2 \pi r^2$ is the surface area of the disk and $\sigma$ is the 
Stefan-Boltzmann constant.  The effective temperature of the disk as a function
of radius $r$ is therefore
$$
T(r) \simeq
    \bigg({G M_\bullet\dot{M_\bullet}\over{4 \pi \sigma r^3}}\bigg)^{1/4}~.
                                                                   \eqno{(3)}
$$
Parameterizing the above result in terms of the Eddington accretion rate, 
$\dot{M}_{\rm E}\,\equiv\,L_{\rm E}/\epsilon c^2$ = 2.2 $(\epsilon/0.1)^{-1}$ 
($M_\bullet/10^8$ \msun) \msund\ yr$^{-1}$, and the Schwarzschild radius, 
$R_S\,\equiv\,2GM_\bullet/c^2$ = $2.95\times10^{13}$ ($M_\bullet/10^8$ \msun) 
cm, gives
$$
T(r)\,=\,6\times 10^5\,{\rm K}\,\bigg({\dot{M_\bullet}
                                \over{\dot{M}_{\rm E}}}
    \bigg)^{1/4}\, \bigg({M_\bullet\over{10^8~M_{\odot}}}\bigg)^{-1/4}\,\bigg(
    {r\over{R_S}}\bigg)^{-3/4}~. \eqno{(4)}
$$
In other words, the peak of the blackbody spectrum occurs at a frequency of 
$\nu_{\rm max}\,=\,2.8{\ts}kT/h\,\simeq$ 4\e{16} Hz, where $k$ is Boltzmann's 
constant and $h$ is Planck's constant.  This peak is near 100 \AA\ or 0.1 keV.
In fact, the spectra of many AGNs show a broad emission excess at extreme
ultraviolet or soft X-ray wavelengths.  This ``big blue bump'' has often been
identified with the thermal emission from the accretion disk.  A fit to the 
luminosity and the central frequency of the big blue bump gives $M_\bullet$ and
$\dot{M_\bullet}$ but not each separately.  Corrections for disk inclination and
relativistic effects further complicate the analysis.  This method is therefore
model-dependent and provides only approximate masses. Typical values for quasars
are $M_\bullet$ $\simeq$ 10$^8$ -- 10$^{9.5}$ \msund\ and $\dot{M_\bullet}$ 
$\simeq$ 0.1 -- 1 $\dot{M}_{\rm E}$.  Seyfert nuclei appear to have lower 
masses, $M_\bullet$ $\simeq$ 10$^{7.5}$ -- 10$^{8.5}$ \msund, and lower 
accretion rates, $\dot{M_\bullet}$ $\simeq$ 0.01 -- 0.5 $\dot{M}_{\rm E}$.

\vfill\eject
\cl {3.2 Virial Masses from Optical Variability}
\vs

      Surrounding the center at a distance of 0.01 to 1 pc from the black hole 
lies the ``\hbox{broad-line} region'' (BLR).  This is a compact, dense, and 
highly turbulent swarm of gas clouds or filaments.  The clouds are illuminated 
by the AGN's photoionizing continuum radiation and reprocess it into emission 
lines that are broadened to velocities of several thousand km s$^{-1}$ by the 
strong gravitational field of the black hole.  Then
$$
      M_\bullet = \eta\, {{v^2 r_{\rm BLR}}\over{G}}~, \eqno{(5)}
$$
where $\eta \simeq 1$ to 3 depends on the kinematic model adopted, $v$ is the
velocity dispersion of the gas as reflected in the widths of the emission lines,
and $r_{\rm BLR}$ is the radius of the BLR.  The latter can be estimated by 
``reverberation mapping,'' as follows.  The photoionizing continuum of an AGN
typically varies on timescales of days to months.  In response, the emission 
lines vary also, but with a time delay that corresponds to the light travel time
between the continuum source and the line-emitting gas.  By monitoring the 
variations in the continuum and the emission lines in an individual object, 
reverberation mapping provides information on the size of the BLR. These studies
also suppport the assumption that the line widths come predominantly from bound
orbital motions.  Applying Equation (5) suggests that Seyfert nuclei are powered
by black holes with masses $M_\bullet \sim 10^7$ to 10$^8$ \msund, while quasar
engines are more massive, with $M_\bullet \sim 10^8$ to 10$^9$ \msund.  Since 
quasars also live in more massive host galaxies, this supports the emerging 
correlation (see the following article) between BH mass and the mass of the 
elliptical-galaxy-like part of the host galaxy.

\vs
\cl {3.3 X-Ray Variability}
\vs

      Active galactic nuclei vary most conspicuously in hard X-rays (2 -- 10 
keV).  One might hope to use the variability timescale to constrain the size of
the X-ray emitting region and hence to estimate the central mass.  However, no
simple pattern of variability emerges, and defining a meaningful timescale is
ambiguous.  One approach uses the ``fastest doubling time,'' $\Delta t$, to 
establish a maximum source size $R \simeq c \Delta t$.  High-energy photons 
presumably come from the hot, inner regions of the accretion disk or in an 
overlying hot corona.  For example, if $R \simeq 5\,R_S$, as deduced in some 
models, we obtain an upper limit to the mass, $M_\bullet$ \lapprox
\ts$(c^3/10G)\, \Delta t$ $\sim$ $10^4\,\Delta t$ \msun ($\Delta t$ in s).
Masses estimated in this way are generally consistent with those obtained from
other virial arguments, but they are considerably less robust because 
of uncertainties in associating the variability timescale with a source size.  
For example, the x-ray intensity variations could originate from localized 
``hotspots'' in the accretion flow.

     X-ray reverberation mapping may in the future be a more powerful tool.  The
iron K$\alpha$ line is widely believed to be produced by reprocessing of the 
hard X-ray continuum by the accretion disk.  The strikingly large width and 
skewness of the line profiles (Figure 1), now routinely detected with 
{\it ASCA\/}, reflect the plasma bulk motion within 10 -- 100 gravitational 
radii of the center.  The temporal response of the line strength and line 
profile depends on a number of factors that, in principle, can be modeled 
theoretically; these include the geometry of the X-ray source, the structure 
of the disk, and the assumed (Schwarzschild or Kerr) metric of the black 
hole.  Time-resolved X-ray spectroscopy should become feasible with the 
{\it X-ray Multi-Mirror Mission\/} ({\it XMM\/}) in the near future.  We can 
then look forward to constraints both on the masses and the spins of BHs.

\vfill

\vs
\cl {4.~SUMMARY AND PROSPECTUS}
\vs

      The black hole model for AGN activity has been successful and popular for
over three decades.  It has withstood the test of time not -- at least until 
recently -- because the empirical evidence for BHs has been overwhelming but
because the alternatives are so implausible.  Now progress has advanced on 
several fronts.  The refurbished {\it HST\/} has greatly strengthened the 
evidence, already growing from ground-based observations, that supermassive dark
objects live at the centers of most galaxies.  The pace of discoveries is
accelerating.  The dark objects have exactly the range of masses that we need to
explain AGN engines, but we have had no proof that they must be black holes.
Then radio interferometry revealed the spectacular maser disk in NGC 4258.  For
its rotation curve to be as accurately Keplerian as we observe, the central mass
must be confined to an astonishingly tiny volume.  The inferred density of the 
central object is so high that astrophysically plausible alternatives can be 
excluded; a BH is the best explanation.  The same conclusion has been reached 
for the BH candidate at the center of our Galaxy.  This is a major conceptual
breakthrough.  

      In addition, {\it ASCA\/} has demonstrated that many AGNs show iron
emission lines with relativistically broadened profiles.  This is arguably the
best evidence for the strong gravitational field of a black hole.  One of the 
most interesting prospects for the future is time-resolved X-ray spectroscopy, 
because hot gas probes closest to an accreting black hole. 

      Finally, the AGN paradigm can be turned inside-out to give what may prove
to be the most direct argument for black holes.  BHs were ``invented'' to 
explain nuclear activity in galaxies.  In recent years, an ironic situation has
developed: some BH candidates are too {\it inactive\/} for the amount of 
matter that we believe they are accreting.  The same is true of some 
stellar-mass black hole candidates that accrete gas from evolving companion 
stars.  A number of researchers recently have developed a theory of 
``advection-dominated accretion'' in which the accretion disk cannot radiate 
most of its energy before it reaches $R_S$ either because it is optically thick 
or because it is too thin to cool.  Unless most of the inflowing material 
ultimately escapes through an outflow, a possibility being explored, the 
only way to make the accretion energy disappear is to ensure that the 
accreting body does not have a hard surface.  That is, the {\it inactivity\/} 
of well-fed nuclear engines may be evidence that they have event horizons.  
Finding event horizons would be definitive proof that AGN engines are black 
holes.

\vs
\cl{5.~SUGGESTIONS FOR FURTHER READING}
\vs

\frenchspacing
\parindent=8pt
\def\nhi{\noindent \hangindent=8pt}

\nhi$\bullet$
Initial debate concerning the physical nature of quasars is summarized in 

\pp
Burbidge, G., \& Burbidge, E.~M., {\it Quasi-Stellar Objects\/} (San 
Francisco: Freeman) (1967)
\bigskip

\nhi$\bullet$
The three key historical papers that originated the BH hypothesis are

\pp
Salpeter, E.~E. \apjo {\bf 140}, 796 (1964)

\pp
Zel'dovich, Ya.~B., \& Novikov, I.~D. {\it Sov. Phys. Dokl.} {\bf 158}, 811 
(1964)

\pp
Lynden-Bell, D. \nato {\bf 223}, 690 (1969)

\nhi$\bullet$ The argument for ``gravity power'' was further developed in

\pp
Lynden-Bell, D. {\it Physics Scripta\/} {\bf 17}, 185 (1978)
\bigskip

\nhi$\bullet$
Textbook style discussions of AGN physics can be found in 

\pp
{\it Active Galactic Nuclei, Saas-Fee Course 20\/}, 
ed. T.~J.-L. Courvoisier \& M. Mayor (Berlin: Springer) (1990)

\pp 
Peterson, B.~M., {\it An Introduction to Active Galactic Nuclei\/} 
(Cambridge: Cambridge University Press) (1997)
\bigskip

\nhi$\bullet$
The BH paradigm is covered at a more technical level in the following review 
articles:

\pp
Rees, M.~J. \annrevo {\bf 22}, 471 (1984)

\pp
Begelman, M. C., Blandford, R. D., \& Rees, M. J. {\it Rev. Mod. Phys.\/} {\bf
56}, 255 (1984)

\pp
Blandford, R.~D., in {\it Active Galactic Nuclei, Saas-Fee Course 
20\/}, ed. T.~J.-L. Courvoisier \& M. Mayor (Berlin: Springer), 161 (1990)
\bigskip

\nhi$\bullet$
The search for BHs is reviewed in

\pp
Kormendy, J., \& Richstone, D. \annrevo {\bf 33}, 581 (1995)

\pp
Richstone, D., \etal \nato {\bf 395}, A14 (1998)
\bigskip

\nhi$\bullet$
The starburst theory for the origin of AGNs has been developed by 

\pp
Terlevich, R., Tenorio-Tagle, G., Franco, J., \& Melnick, J. \mnraso {\bf 
255}, 713 (1992)

\pp
Terlevich, R., Tenorio-Tagle, G., Rozyczka, M., Franco, J., \& Melnick, J.
\mnraso {\bf 272}, 198 (1995)
\bigskip

\nhi$\bullet$
The following conference proceedings explicitly focus on the observations 
and interpretation of the more ``garden variety'' low-luminosity AGNs:

\pp
Eracleous, M., Koratkar, A.~P., Leitherer, C., \& Ho, L.~C., eds., 
{\it The Physics of LINERs in View of Recent Observations\/}
(San Francisco: Astronomical Society of the Pacific) (1996)

\pp
Schmitt, H.~R., Kinney, A.~L., \& Ho, L.~C., eds., {\it The AGN/Normal Galaxy 
Connection} ({\it Advances in Space Research,} {\bf 23 (5-6)}) (Oxford: 
Elsevier Science Ltd.) (1999)
\bigskip

\nhi$\bullet$
Readers interested in a full treatment of the techniques of 
reverberation mapping should consult

\pp
Blandford, R.~D., \& McKee, C.~F. \apjo {\bf 255}, 419 (1982)

\pp
Peterson, B.~M. \paspo {\bf 105}, 247 (1993)
\bigskip

\nhi$\bullet$
Explicit application of reverberation mapping results to derive masses 
of AGNs was done by 

\pp
Ho, L.~C., in {\it Observational Evidence for Black Holes in the 
Universe\/}, ed. S.~K. Chakrabarti (Dordrecht: Kluwer), 157 (1998)

\pp
Laor, A. \aplo {\bf 505}, L83 (1998)
\bigskip

%\nhi$\bullet$
%The following paper reviews nuclear gas disks discovered with the
%{\it Hubble Space Telescope\/}:

%\pp
%Ford, H.~C., Tsvetanov, Z.~I., Ferrarese, L. \& Jaffe, W., in {\it IAU 
%Symposium 184, The Central Regions of the Galaxy and Galaxies}, ed. Y. Sofue
%(Dordrecht: Kluwer), 377 (1998)
%\bigskip

\nhi$\bullet$
Mass determinations using optical emission-line rotation curves include

%\pp
%Ferrarese, L., Ford, H.~C., \& Jaffe, W. \apjo {\bf 470}, 444 (1996)
%This is a terrible paper; we can tell them about much better ones.

\pp
Harms, R. J., \etal \aplo {\bf 435}, L35 (1994)

\pp
Macchetto, F., Marconi, A., Axon, D.~J., Capetti, A., Sparks, W.~B., \&
Crane, P. \apjo {\bf 489}, 579 (1997)

\pp
Bower, G.~A., \etal \aplo {\bf 492}, L111 (1998)
\bigskip

%\vfill\eject

\nhi$\bullet$
The water maser observations of NGC 4258 are described in

\pp
Watson, W.~D., \& Wallin, B.~K. \aplo {\bf 432}, L35 (1994)

\pp
Miyoshi, M., Moran, J., Herrnstein, J., Greenhill, L., Nakai, N., Diamond, P.,
\& Inoue, M. \nato {\bf 373}, 127 (1995)
\bigskip

\nhi$\bullet$
Arguments against compact dark star clusters in NGC 4258 and the Galaxy are 
presented in

\pp
Maoz, E. \aplo {\bf 494}, L181 (1998)
\bigskip

\nhi$\bullet$
These papers discuss the derivation of $M_\bullet$ and $\dot{M_\bullet}$ by
fitting spectra with accretion disk models:

\pp
Wandel, A., \& Petrosian, V. \aplo {\bf 329}, L11 (1988)

\pp
Laor, A. \mnraso {\bf 246}, 369 (1990)
\bigskip

\nhi$\bullet$
Attempts to derive masses using X-ray variability have been made by

\pp
Wandel, A., \& Mushotzky, R.~F. \aplo {\bf 306}, L61 (1986)
\bigskip

\nhi$\bullet$
The prediction, discovery, and routine detection of broad Fe K\al\ emission
lines are described, respectively, in

\pp
Fabian, A.~C., Rees, M.~J., Stella, L., \& White, N.~E. \mnraso
{\bf 238}, 729 (1989)

\pp
Tanaka, Y., \etal \nato {\bf 375}, 659 (1995)

\pp
Nandra, K., George, I.~M., Mushotzky, R.~F., Turner, T.~J., \& Yaqoob, T.
\apjo {\bf 477}, 602 (1997)
\bigskip

\nhi$\bullet$
Prospects for X-ray reverberation mapping are foreseen in 

\pp
Stella, L. \nato {\bf 344}, 747 (1990)

\pp
Reynolds, C.~S., Young, A.~J., Begelman, M.~C., \& Fabian, A.~C. \apjo
{\bf 514}, 164 (1999)
\bigskip

\nhi$\bullet$
Advection-dominated accretion is reviewed in

%\pp
%Menou, K., Quataert, E., \& Narayan, R., in {\it Proceedings of the Eighth
%Marcel Grossmann Meeting on General Relativity\/} (Jerusalem, June 1997),
%in press

\pp
Narayan, R., Mahadevan, R., \& Quataert, E., in {\it The Theory of Black Hole
Accretion Discs\/}, ed. M. A. Abramowicz, G. Bj\"{o}rnsson, \& J. E. Pringle
(Cambridge: Cambridge University Press), 148 (1998)

\pp
Mineshige, S., \& Manmoto, T., {\it Advances in Space Research}, 
{\bf 23 (5-6)}, 1065 (1999)

\pp
Blandford, R.~D., \& Begelman, M.~C. \mnraso {\bf 303}, L1 (1999)

\end

\nhi$\bullet$
The case for and against the binary black hole interpretation of radio 
galaxies with double-peaked emission lines are made, respectively, by

\pp
Gaskell, C.~M., in {\it Jets from Stars and Galaxies\/}, ed. W. Kundt (Berlin:
Springer), 165 (1996)

\pp
Eracleous, M., Halpern, J.~P., Gilbert, A.~M., Newman, J.~A. \& Filippenko,
A.~V. \apjo {\bf 490}, 216 (1997)
\bigskip